\documentclass[aps,prb,twocolumn,showpacs,preprintnumbers,amsmath,amssymb,superscriptaddress]{revtex4}
\usepackage{graphicx}
\usepackage{tabularx}
\usepackage{color}
\usepackage{xspace}

\newcommand{\msr}{$\mu$SR\xspace }

\begin{document}
\preprint{PREPRINT (\today)}

\title{$\mu$SR investigation of magnetism and magnetoelectric coupling in  Cu$_2$OSeO$_3$}

\author{A.~Maisuradze}\email{alexander.m@physik.uzh.ch}
\affiliation{Physik-Institut der Universit\"{a}t Z\"{u}rich, Winterthurerstrasse 190, CH-8057 Z\"{u}rich, Switzerland}
\affiliation{Laboratory for Muon Spin Spectroscopy, Paul Scherrer Institut, CH-5232 Villigen PSI, Switzerland}
\author{Z.~Guguchia}
\affiliation{Physik-Institut der Universit\"{a}t Z\"{u}rich, Winterthurerstrasse 190, CH-8057 Z\"{u}rich, Switzerland}
\author{B.~Graneli}
\affiliation{Physik-Institut der Universit\"{a}t Z\"{u}rich, Winterthurerstrasse 190, CH-8057 Z\"{u}rich, Switzerland}
\affiliation{Institute of Theoretical Physics, ETH H\"onggerberg, CH-8093 Z\"urich, Switzerland}
\author{H. M.  R\o{}nnow}
\affiliation{Institute of Condensed Matter Physics, \'{E}cole Polytechnique F\'{e}d\'{e}rale de Lausanne (EPFL),
CH-1015 Lausanne, Switzerland}
\author{H. Berger}
\affiliation{Institute of Condensed Matter Physics, \'{E}cole Polytechnique F\'{e}d\'{e}rale de Lausanne (EPFL),
CH-1015 Lausanne, Switzerland}
\author{H.~Keller}
\affiliation{Physik-Institut der Universit\"{a}t Z\"{u}rich, Winterthurerstrasse 190, CH-8057 Z\"{u}rich, Switzerland}

\begin{abstract}
A detailed zero and transverse field (ZF\&TF) muon spin rotation (\msr) investigation
of magnetism and the magneto-electric coupling in Cu$_2$OSeO$_3$ is reported.
An internal magnetic field $B_\mathrm{int}(T=0) = 85.37(25)$ mT was
found, in agreement with a ferrimagnetic state below $T_{\rm c} = 57.0(1)$ K.
The temperature dependence of the magnetic order parameter is well described by the
relation $B_{\rm int} = B(0)(1-(T/T_{\rm c})^2)^{\tilde{\beta}}$ with an
effective exponent $\tilde{\beta}\simeq 0.39(1)$
which is close to the critical exponent $\beta \simeq 1/3$ for a three dimensional (3D) magnetic system.
Just above $T_{\rm c}$ the muon relaxation rate follows the power low
$\lambda(T) \propto (T/T_{\rm c}-1)^{-\tilde{\omega}}$ with ${\tilde{\omega}}=1.06(9)$,
which is characteristic for 3D ferromagnets.
Measurements of $B_{\rm int}(T)$ with and without an applied electrostatic field
$E=1.66\times10^5$ V/m suggest a possible electric field effect of magnitude
$\Delta B_V =  B_V(0{\rm V})-B_V(500{\rm V}) = - 0.4(4)$~mT.

\end{abstract}

\maketitle

\section{Introduction}

Much attention has been directed toward multiferroic and magnetoelectric materials in
recent years.\cite{Eerenstein, Spaldin05, Cheong07, Fiebig05}
The coupling between magnetic and electric parameters increases the degrees
of freedom of the ordered ground state, making these materials
good candidates for the study of new phenomena in highly correlated electronic systems.
Strong magnetoelectric coupling is rather rare in the solid state, since usual microscopic mechanisms
for magnetic and electric polarization are mutually exclusive. Magnetism requires strong
exchange interactions related to a strong hybridization of the transition ion electrons leading to
conductivity. Conductivity, on the other hand, is inconsistent with the presence of an electric
polarization in a sample.\cite{Hill00}
It is therefore of particular importance to unravel the mechanisms behind magnetoelectric coupling.
A number of atomic mechanisms have been proposed in order to explain the magnetoelectric coupling.\cite{Fiebig05,Gehring94,Sergienko06,Katsura05}
Considering spatial and time inversion
symmetry for the magnetic (${\bf M}$) and the electric (${\bf P}$) polarization, it was concluded that
linear magnetoelectric coupling is only possible when both vectors vary in space and time.\cite{Cheong07}
On the other hand, the importance of frustration effects
in magnetoelectrics for the static polarizations of {\bf P} and {\bf M} was stressed for
nonlinear coupling mechanisms.\cite{Cheong07}
The presence of large magnetic and electric polarizations is an important condition for
strong magnetoelectric coupling, making ferro- or ferrimagnetic materials favorable candidates.\cite{Fiebig05}

The ferrimagnetic magnetoelectric compound Cu$_2$OSeO$_3$ was recently discovered,\cite{Bos08,Effenberger86}
and single crystals were successfully grown soon after.\cite{Belesi10}
The compound is piezoelectric and undergoes a ferrimagnetic transition below 60 K, exhibiting magnetoelectric
coupling as revealed by magneto-capacitance studies on a polycrystalline sample.\cite{Bos08}
An abrupt change of the dielectric constant below the ferrimagnetic transition was later confirmed by
infrared studies.\cite{Miller10,Gnezdilov10}
At present the nature of the magnetoelectric coupling is unknown, since the most common mechanisms,
involving magnetostriction and piezoelectric effects via lattice
distortions are excluded. Neither X-ray diffraction (XRD)\cite{Bos08} nor
nuclear magnetic resonance (NMR)\cite{Belesi10} studies revealed any lattice
anomaly below the N\'{e}el temperature.

The positive muon  is a very sensitive microscopic probe for
studying magnetic properties in zero as well as in an applied electric field.\cite{Blundell99muSR}
Following the pioneering works of Eschchenko {\it et al.}\cite{Eshchenko99} and Lewtas {\it et al.}\cite{Lewtas10}
we implemented a setup with alternating electric fields, and performed a muon spin rotation/relaxation (\msr) investigation of magnetism and magnetoelectric coupling in Cu$_2$OSeO$_3$.
The temperature dependence of the  internal magnetic field $B_{\rm int}$ was investigated
below $T_{\rm c} = 57.0(1)$ K,  and the relaxation rate was studied above $T_{\rm c}$.
Zero field \msr measurements of the internal field distribution with and without an applied electric field
$E=1.66\times10^5$ V/m indicate a small electric field effect on the internal magnetic
field: $\Delta B_V =  B_V(0{\rm V})-B_V(500{\rm V}) = - 0.4(4)$ mT.

The paper is organized as follows: The sample preparation and the details of experimental setup are
described in Sec. \ref{sec:expDet}. In Sec. \ref{sec:analysis} we describe the model
used for the analysis of the $\mu$SR data and the relation of the measured $\mu$SR spectrum to
the lattice and magnetic structure of the sample. In Sec. \ref{sec:ResAndDisc} we present
and discuss the obtained results. Conclusion is given in Sec. \ref{sec:conclusions}.


\section{Experimental details}\label{sec:expDet}

A high quality single crystal of Cu$_2$OSeO$_3$ of approximate size 7$\times$8$\times$3 mm$^3$ was prepared in a manner
described elsewhere.\cite{Belesi10} The zero field (ZF) and transverse field (TF) \msr experiments were
performed at the $\pi$E3 beam line at the Paul Scherrer Institute (Villigen, Switzerland).
The crystal structure is cubic with symmetry (P$2_13$).\cite{Bos08,Effenberger86}
The sample was aligned with its (100) direction
parallel to the incident muon beam. The spin vector of the muon was oriented approximately with an angle of 60 degrees
with respect to the momentum vector. The asymmetry time spectra were monitored in the "Forward", "Backward",
"Up", and "Down" (FBUD) positron detectors.\cite{Blundell99muSR}
Typical statistics was 40 to 50$\times10^6$ positron events in the FBUD detectors for a spectrum
with a 5\,$\mu$s time window. The switched electric field $E$ was
applied along the (100) direction of the 3 mm thick crystal.
The crystal was mounted between two Cu electrodes: A thin (50$\mu$m)
Cu metal foil was used as the positive electrode, and the negative electrode was soldered to the
sample holder and electrically connected to ground.
The applied voltage was switched at a rate of 100 Hz for two reasons: (i) to avoid accumulation of
muon created charge in the vicinity of the sample-to-electrode contact surface, which might offset and
even cancel the applied field $E$, and (ii) to provide a well-defined consecutive reference in order to reduce artefacts
related to any slight variation in temperature or applied magnetic field with time.
All the positron events registered by the FBUD detectors were stored alternatively in the first block of
four histograms when the electric field was off, and the second block of four histograms when
the electric field was on.
Measurements were performed for two different electric fields: $E = 500/3$ V/mm and $E=800/3$ V/mm
(i.e. 500 or 800 V applied on 3 mm thick sample).

\section{Analysis and models}\label{sec:analysis}

ZF \msr allows to determine internal magnetic fields at the position in the lattice where the muons stop.
For a polycrystalline sample with a given static magnetic field $B$ at the stopping site
of the muon, the muon depolarization function\cite{Blundell99muSR} may be expressed as:
$G_B(t) = \frac{1}{3} + \frac{2}{3}\cos(\gamma_\mu B t)$,
where $\gamma_\mu = 2\pi\times 135.53$~MHz/T
is the gyro-magnetic ratio of muon.  $G_B(t)$ consists of two parts: a constant fraction of 1/3,
and a fraction $2/3$ oscillating with the frequency $\omega = \gamma_\mu B$.
For a given magnetic field distribution $P(B)$, the muon depolarization function at the muon site is:\cite{Caretta09}
\begin{equation}\label{ea:PBasy}
G_P(t) = \frac{1}{3} + \frac{2}{3}\int_0^\infty P(B)\,\cos(\gamma_\mu B t)\, dB.
\end{equation}
The function $P(B)$ contains information on the magnetic structure and the spatial magnetic field
distribution of the sample,
but also the effect of a static electric field $E$ on the local magnetic fields in the sample.
%

\begin{figure}[tb]
\includegraphics[width=0.97\linewidth]{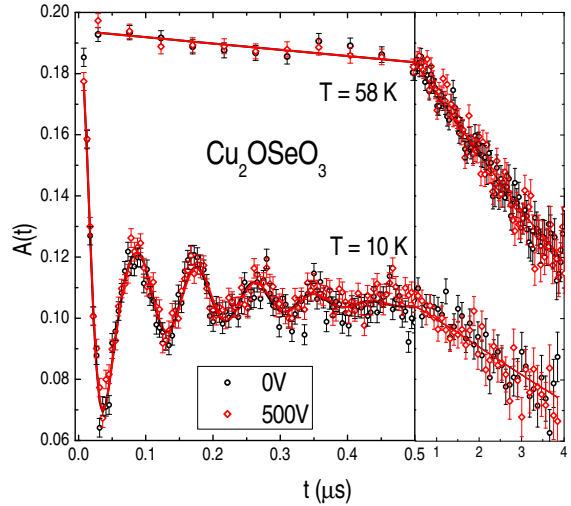}
\caption{ (Color online) \msr time spectra of Cu$_2$OSeO$_3$ at 10 and 58 K
in ZF with ($\lozenge$) and without ($\circ$) applied electrical field $E=500/3$ V/mm. Solid lines are
fits to the data using Eqs. (\ref{eq:PBasyCu}) and (\ref{eq:PB}). }
\label{fig:Asy}
\end{figure}
\begin{figure}[h]
\includegraphics[width=0.97\linewidth]{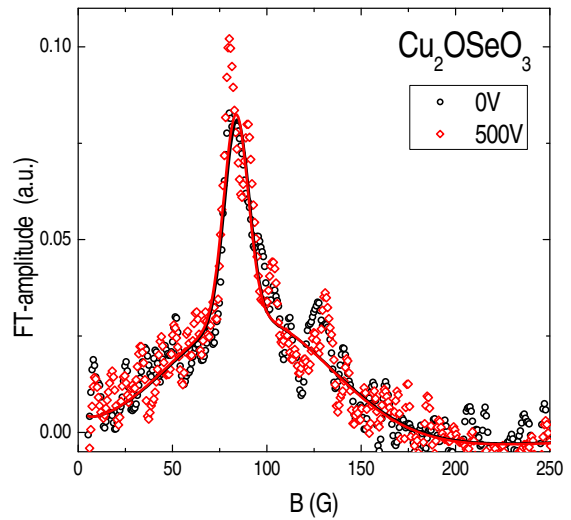}
\caption{ (Color online) Fourier transform of the oscillating part
of the asymmetry spectra shown in Fig. \ref{fig:Asy} (black circles: $E=0$; red diamonds: $E=500/3$ V/mm).
The solid lines are the FT of the corresponding solid curves
in Fig. \ref{fig:Asy}}\label{fig:FT}
\end{figure}

Due to an additional isotropic dynamic muon relaxation and the presence of a Cu background
signal, the ZF depolarization function for Cu$_2$OSeO$_3$ is best described by the following equation:
\begin{align}\label{eq:PBasyCu}
A(t) = &A_S\cdot\left( \frac{1}{3} + \frac{2}{3}\int_0^\infty
P(B)\,\cos(\gamma_\mu B t + \phi)\, dB \right)e^{-\lambda t}+\nonumber  \\
   +& A_{Bg}\cdot G_{KT}(\sigma_{Bg} t)\cdot e^{-\lambda_{Bg} t}.
\end{align}
Here $A_S$ is proportional to the fraction of muons stopping in the sample,
and $A_{Bg}$ is proportional to the fraction of muons stopping in the Cu sample holder.
The parameters $\lambda_{Bg}$ and $\sigma_{Bg}$ describe the temperature independent
muon depolarization in Cu, while $\lambda$ corresponds to the dynamic
muon relaxation in the sample. $G_{KT}(\sigma_{Bg} t)$ denotes the Gaussian Kubo-Toyabe depolarization
function.\cite{Kubo97} Note that the phase $\phi  = 0$ for zero field.
As for a powder sample, we observe also for the present single crystal sample a static fraction of 1/3 for the
depolarization function and an oscillating fraction of 2/3.\cite{Blundell99muSR} In the case of a single
crystal sample, though, these fractions are due to the formation of magnetic domains with a random spatial distribution
which in fact corresponds to the situation for a polycrystalline sample.
The best fit of $P(B)$ to the experimental data with a minimal set of parameters was obtained with a sum of two Gaussians:
\begin{equation}\label{eq:PB}
P(B) = \sum_{i=1,2} \frac{F_{S}^i}{\sqrt{2\pi}\sigma_i/\gamma_\mu}\cdot\exp\left[-\frac{1}{2}\left(\frac{B-B_i}{\sigma_i/\gamma_\mu}\right)^2 \right].
\end{equation}
Here $F_{S}^1$ and $F_{S}^2$ are the fractions of the two Gaussians with mean fields
$B_1$ and $B_2$ and standard deviations $\sigma_1/\gamma_\mu$ and $\sigma_2/\gamma_\mu$.
Analysis of the data measured with highest statistics ($100\times 10^6$ positron events) at 10 K
with Eqs. (\ref{eq:PBasyCu}) and (\ref{eq:PB})
[see Figs. (\ref{fig:Asy}) and (\ref{fig:FT})] yields in $F_{S}^1=0.18(2)$,  $F_{S}^2=0.82(2)$,
and  $B_2/B_1 = 1.07(2)$ [note that $F_{S}^1+F_{S}^2=1$]. The fit were performed by keeping all the sample
parameters ($A_S$, $\sigma_1$, $\sigma_2$, $\lambda$, $F_{S}^1$, $F_{S}^2$, and $B_2/B_1$) and the
background parameters ($A_{Bg}$, $\sigma_{Bg}$, and $\lambda_{Bg}$)
the same, while  $B_1(0 {\rm V})$ and $B_1(500 {\rm V})$ (data without
and with the applied electric field) were free parameters.
The temperature independent sample asymmetry $A_S=0.144$ and the background parameters
$A_{Bg} = 0.05$, $\sigma_{Bg}=0.14(2)$ $\mu$s$^{-1}$, and $\lambda_{Bg}=0.11(2)$ $\mu$s$^{-1}$
were determined from the global fit of the whole temperature dependence,
while the total initial asymmetry $A_S + A_{BG} = 0.194$ was determined above $T_{\rm c}$.
Further more, all the background parameters and the following sample parameters: $F_{S}^1$, $F_{S}^2$, $B_2/B_1$,
were kept as temperature independent.

Figure \ref{fig:Asy} shows the asymmetry time spectra at 10\,K and 58\,K (below and above the magnetic transition)
with and without electrostatic field $E$. The corresponding fits
using Eqs. (\ref{eq:PBasyCu}) and (\ref{eq:PB}) are represented by the lines.
Figure \ref{fig:FT} shows the Fourier transform (FT) amplitudes of the oscillating part of the
\msr spectra and the fitted curves shown in Fig.~\ref{fig:Asy}.
For a small relaxation rate $\lambda$ these FT amplitudes
represent the magnetic field distribution $P(B)$ given by Eq.~(\ref{eq:PB}).
Note that $P(B)$ consisting of two Gaussians [Eq.~(\ref{eq:PB})]
describes the basic features of the $\mu$SR spectrum quite well,
and that the FT amplitudes for $E=0$ and $E=500/3$~V/mm are almost identical.

\begin{figure}[!htb]
\includegraphics[width=0.95\linewidth]{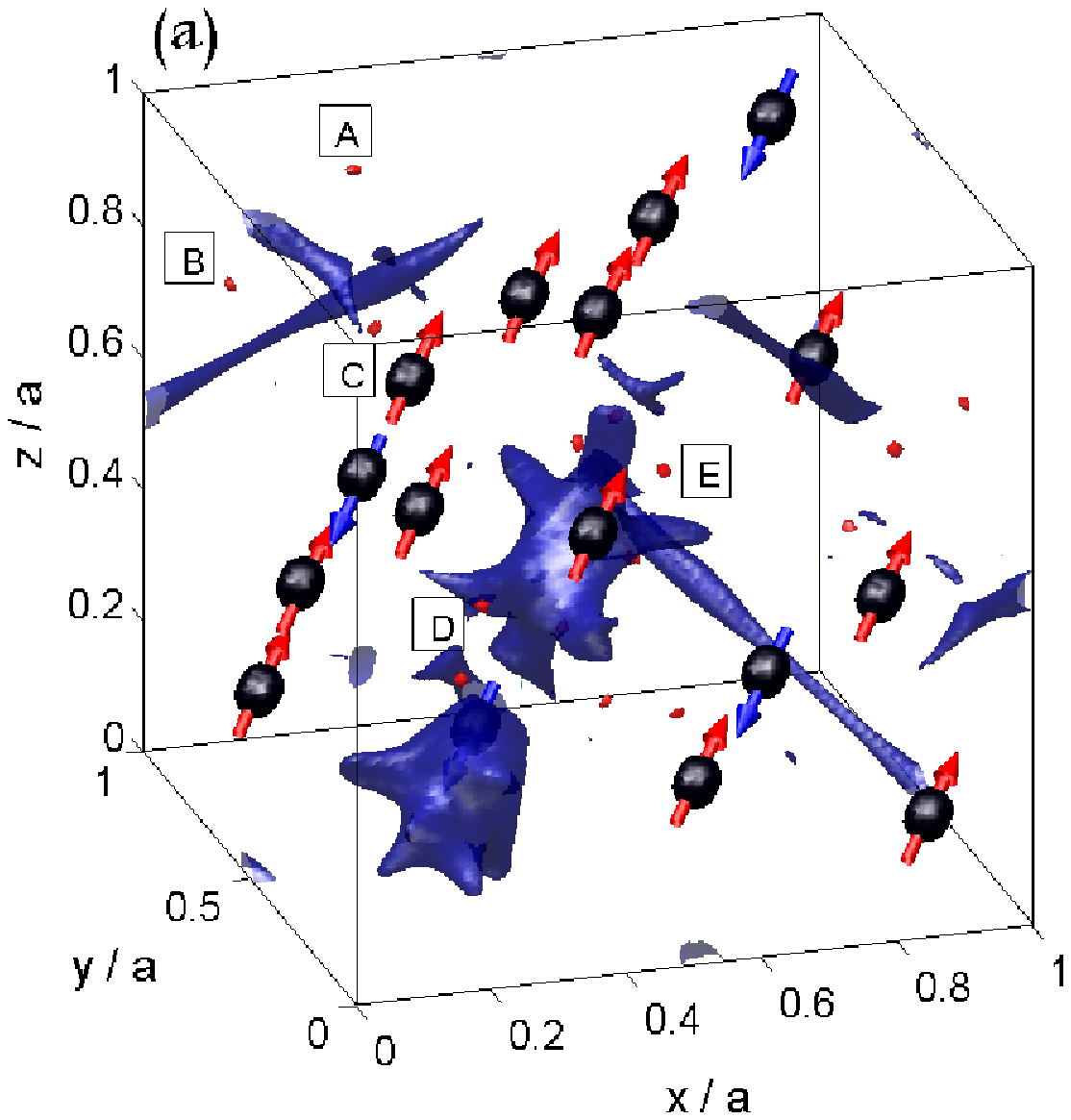}
\includegraphics[width=0.95\linewidth]{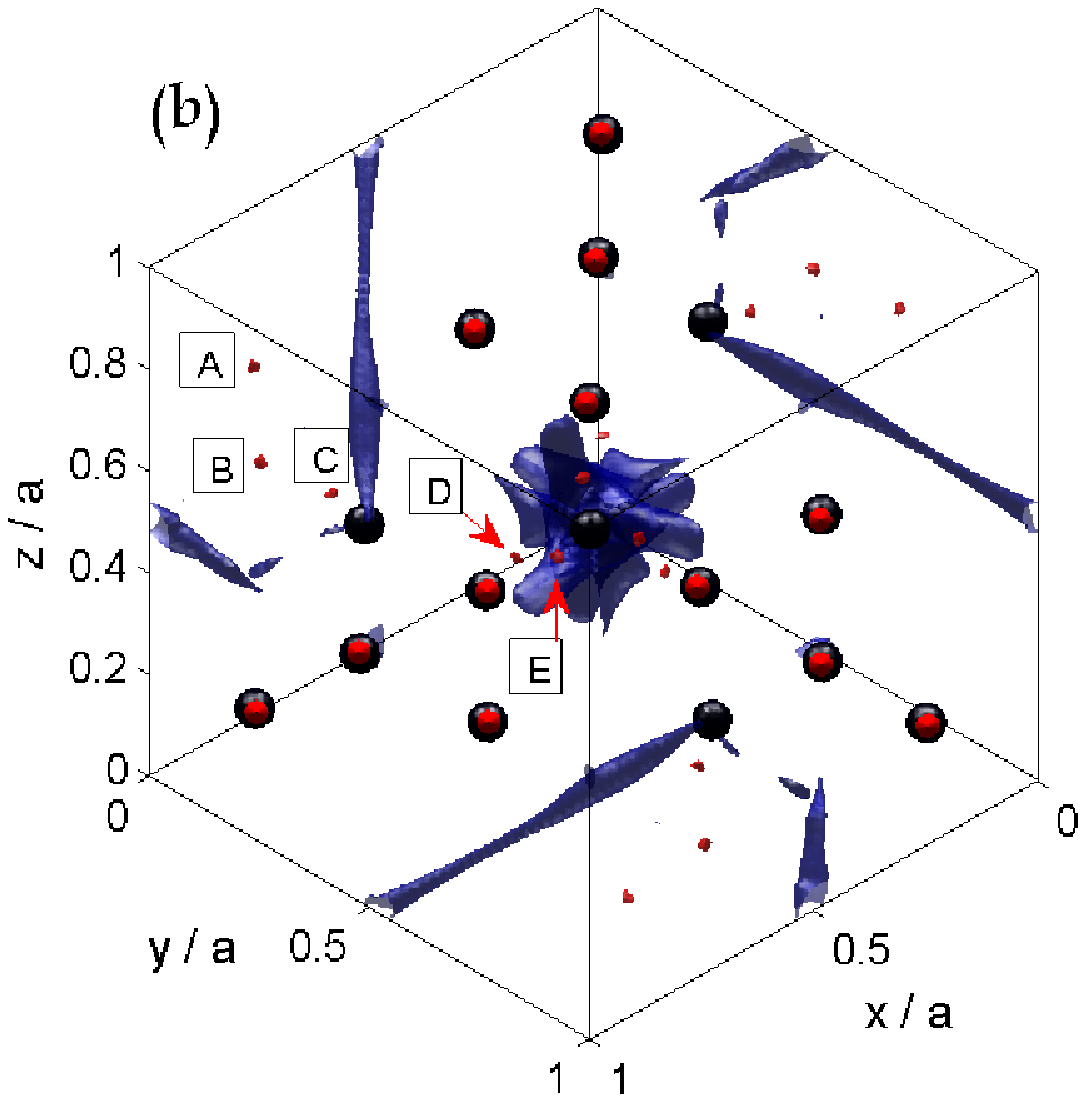}
\caption{ (Color online) Magnetic structure and dipolar fields in Cu$_2$OSeO$_3$ (a). The black
spherical surfaces represent a field strength of 20 T (i.e. Cu$^{2+}$ ions). The arrows
indicate the magnetic moments of Cu$^{2+}$. Blue surfaces represent internal fields of 10 mT (for 0.5$\mu_B$ per Cu ion).
The red spots A, B, C, D, and E are the muon stopping sites.
Panel (b) shows the same as (a) but in the (111) direction.}\label{fig:CrystSrtuct}
\end{figure}

\begin{figure}[tb]
\includegraphics[width=0.95\linewidth]{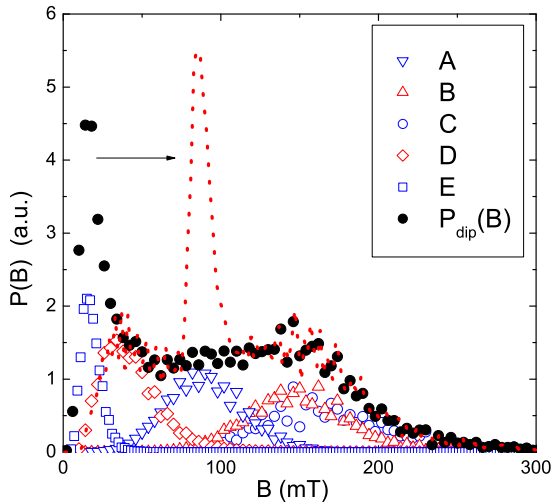}
\caption{ (Color online)
Probability distribution of dipolar fields $P_\mathrm{dip}(B)$ with equal weights for sites
A, B, C, D, and E ($\bullet$).
The contributions of the individual sites A, B, C, D, and E to $P_{\rm dip}(B)$ are
represented by the symbols, $\triangledown$, $\triangle$, $\circ$, $\lozenge$, and $\square$,
respectively. The width of each curve is proportional to the magnetic field gradient at the
corresponding site. The dotted line shows the $P_\mathrm{dip}(B)$ with an additional contact field of 70 mT
at site E. }\label{fig:PBmodel}
\end{figure}

Before presenting the experimental results, we describe below the relation between the measured $\mu$SR
spectra and the magnetic structure of Cu$_2$OSeO$_3$.
The crystal symmetry of Cu$_2$OSeO$_3$ is cubic with a lattice constant $a=8.9111$~\AA\xspace and
spatial group symmetry P$2_13$.\cite{Bos08}
The lattice structure is thus quite complex with 32 oxygen ions in the unit cell,
suggesting quite a large number of possible muon stopping sites.
Generally, the positive muon stops at a high symmetry interstitial site of the lattice,
close to  negatively charged ions (in this case O$^{2-}$). In some cases the muon
may form weak bonds with oxygen.\cite{Blundell99muSR}
The function $P(B)$ describes the muon site-weighted distribution of internal fields in the sample.

In order to find the muon stopping sites, an analysis of the electrostatic potential 
was performed. The potential at position ${\bf r}$
within the lattice unit cell was approximated using a spherical point charge-like
model (in dimensionless units):
\begin{equation}
V({\bf r}) =  \sum_j\left( \frac{q_j}{R_j} + \frac{V^0_{xc}}{R_j}e^{-2R_j/R^I_j}\right).
\end{equation}
The first term is the Coulomb potential, while the second term is the exchange correlation
potential, which is usually assumed to be proportional to the local charge density
(the expression $\exp(-R_j/R^I_j)$ represents an average radial wave function).\cite{Meier83,Velders89}
Here, $R_j=|{\bf r}-{\bf r}_j|$, and the sum is taken over ion coordinates ${\bf r}_j$ within a
cluster of $4\times 4\times 4$ unit cells. The charges $q_j$ and the ionic radii
$R^I_j$ are: $+2$ and 0.71 \AA, $+4$ and 0.42 \AA, $-2$ and 1.3 \AA\xspace for Cu, Se, and O
ions, respectively (here the elementary charge is unity).
The adjustable parameter $V^0_{xc}$ was chosen as $\simeq +10$.
We found that the coordinates of the potential minima do not appreciably depend on
$V^0_{xc}$ over a broad range of values.
The potential $V({\bf r})$ has five magnetically non-equivalent
minima with nearly equal depth at the following sites:
A = (0.215,0.700,0.970), B = (0.035, 0.720, 0.805), C = (0.195, 0.555, 0.795),
D = (0.275, 0.295, 0.460), and E = (0.635, 0.550, 0.525). The sites A, B, C, D, and E are indicated
as red spots in Fig. \ref{fig:CrystSrtuct}. In addition, there are four local minima with higher
energy and lower probability to be occupied.

The muon probes the vector sum of the internal (dipolar) magnetic field and the contact field
at a particular lattice cite.
The dipolar magnetic field  ${\bf B}({\bf r})$ at position ${\bf r}$
within the lattice unit cell is calculated as follows:\cite{Blundell09DipFld}
\begin{equation}\label{eq:DipFld}
B_\mathrm{dip}^\alpha({\bf r}) = \frac{\mu_0}{4 \pi}\sum_{i,\beta} \frac{m_i^\beta}{R_i^3}
\left( \frac{3R_i^\alpha R_i^\beta}{R_i^2} -\delta^{\alpha \beta} \right) .
\end{equation}
Here ${\bf R}_i = {\bf r} - {\bf r}_i$, $\alpha$ and $\beta$ denote the vector components $x$, $y$, and $z$, ${\bf r}_i$ is the position of $i$-th magnetic ion in the unit cell,
and $m^{\beta}_i$ is the corresponding dipolar moment. The summation is taken over a sufficiently
large Lorentz sphere of radius $R_L$. Beyond the Lorentz sphere, the integration is carried out
over the domain volume. The contribution
to the internal magnetic field from this integral is ${\bf B}' = 4\pi\mu_0 (\frac{1}{3}-{\bf \hat{N}_d}) {\bf M}$,\cite{WhiteBook70}
where $\mu_0 {\bf M}\simeq 66$~mT is the domain magnetization,  and ${\bf \hat{N}_d}$ the demagnetization
tensor determined by the geometry of the domain and the magnetic anisotropy.
For the calculation of the magnetization the lattice constant $a=8.91113$ \AA\xspace
and the magnetic moment of 0.5$\mu_B$ per Cu$^{2+}$ ion
were taken from Ref. \onlinecite{Bos08}.
For a magnetically isotropic spherical domain
${\bf \hat{N}_d} = \frac{1}{3}$.\cite{WhiteBook70,Blundell09DipFld} The field ${\bf B}'$ and the stray
fields due to the neighboring domains average statistically to zero.
They only give rise to an additional broadening of $P(B)$, which is smaller than
the width of the narrow component of $P(B)$ (see Fig. \ref{fig:FT}).
The magnetic structure of Cu$_2$OSeO$_3$ and the spatial magnetic
field distribution calculated with Eq.~(\ref{eq:DipFld})
is shown in Fig.~\ref{fig:CrystSrtuct}.
The probability field distributions for the magnetic structure with equal weights for the
muon sites A, B, C, D, and E are shown in Fig. \ref{fig:PBmodel}. The calculations were performed with
Gaussian sampling around the points A, B, C, D, and E, with a standard deviation $\sigma_L=0.23$~\AA.
Thus, the widths of the curves in Fig. \ref{fig:PBmodel} are proportional to the magnetic field gradients at
the corresponding sites. The total field distribution from all
sites A, B, C, and D has a broad Gaussian-like shape
centered at approximately 100\,mT.
This broad distribution agrees quite well with the experimental distribution (see Fig. \ref{fig:FT}).
The narrow peak calculated for site E is located at about 15\,mT,
in contrast to the experimentally observed peak at around 85\,mT. This discrepancy may be explained
by assuming an additional contact field
of approximately 70\,mT at the muon site E, resulting in a peak position of 85\,mT (the total field is the vector
sum of the dipolar field ${\bf B}_\mathrm{dip}$ and the contact field ${\bf B}_c$). Note that the ratio of the broad and the
narrow signal intensities is about 4, in good agreement with the ratio $F_{S}^2/F_{S}^1=4.5(5)$ obtained
from Eq. (\ref{eq:PB}) above.

\begin{figure}[!tb]
\includegraphics[width=0.97\linewidth]{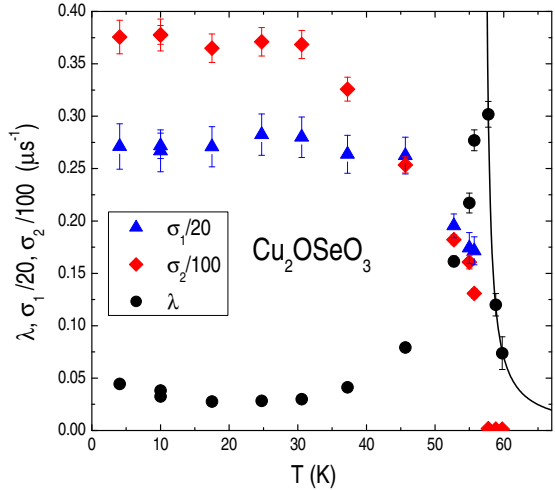}
\caption{ (Color online) Transverse relaxation rates $\sigma_1$ and $\sigma_2$ as well as
longitudinal relaxation rate $\lambda$ of single-crystal Cu$_2$OSeO$_3$ as a function of temperature.
The black solid line represents a fit of the data to the power law
$\lambda(T) = A(T/T_c-1)^{-\tilde{\omega}}$  (see text). }\label{fig:RlxZF}
\end{figure}

\begin{figure}[!tb]
\includegraphics[width=0.97\linewidth]{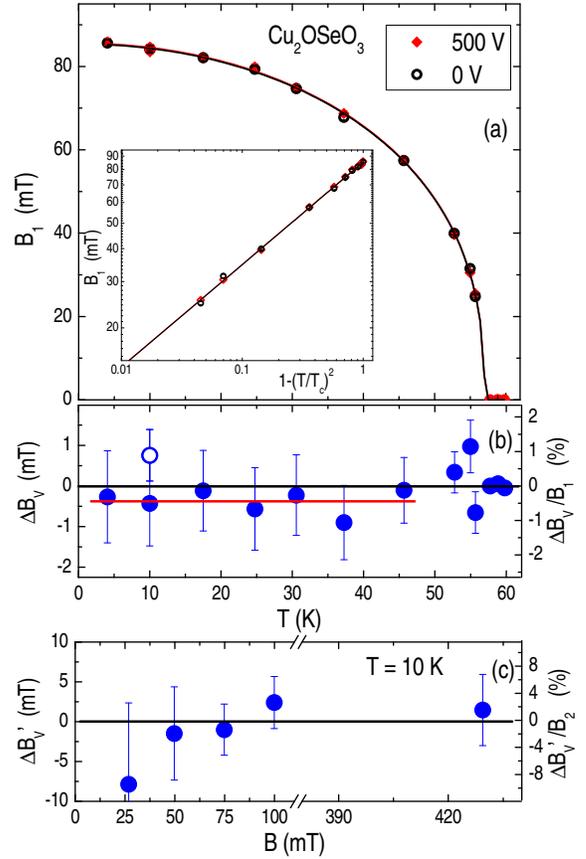}
\caption{ (Color online) (a) Temperature dependence of the mean internal $B_1$ field in single
crystal Cu$_2$OSeO$_3$ for zero and applied electrostatic field $E$. The solid lines are power
law fits to the data with Eq. (\ref{eq:PowerLow}). The insert shows $B_1$ vs $(1-(T/T_{\rm c})^2)$ on a log-log
scale.(b) Electric field shift
$\Delta B_{\rm V} = B_1(0{\rm V})-B_1(500{\rm V})$ as a function of temperature. The solid red line
corresponds to the mean value
of $\overline{\Delta B_{\rm V}}\simeq -0.4(4)$ mT below 50 K. (c) Electric field shift
$\Delta B_{\rm V}' = \hat{B}_2(0{\rm V})-\hat{B}_2(800{\rm V})$ as a function of magnetic field
measured in the TF \msr experiment at 10K.  }\label{fig:BintZF}
\end{figure}

\section{Results and discussion}\label{sec:ResAndDisc}

The temperature dependence of the parameters $\sigma_1$, $\sigma_2$, and $\lambda$
as obtained from  the data analysis by means of Eqs. (\ref{eq:PBasyCu}) and (\ref{eq:PB})
are displayed in  Fig. \ref{fig:RlxZF}.
The parameters $\sigma_1$ and $\sigma_2$ decrease with
increasing temperature and drop to zero at the Curie temperature $T_{\rm c}$ = 57.0(1) K
of the ferrimagnetic transition of Cu$_2$OSeO$_3$. The longitudinal relaxation rate $\lambda$ is
a measure of the internal magnetic field dynamics and can be expressed by the field-field correlation function
$\lambda=\gamma_\mu^2\int_0^\infty <B_\perp(t)B_\perp(0)>dt$.\cite{Pratt07}
Here $B_\perp$ is the field component perpendicular to the muon-spin direction at the muon site.
Brackets denote statistical averages.
In the paramagnetic state close to
$T_{\rm c}$ the relaxation rate $\lambda$ is a measure of the spatial correlation length
$\xi$ of the magnetic order. In the critical state the following relations
hold:  $\lambda \propto \xi^{3/2}$ for ferromagnets and
$\lambda \propto \xi^{1/2}$ for antiferromagnets.\cite{Yaouanc93,Lovesay95EuO,Lovesay95RbMnF}
Note that these relations are only strictly valid in the critical regime very close to $T_{\rm c}$.
However, it was found empirically, that these relations describe experimental data rather well for
EuO, EuS, RbMnF$_3$ up to $\sim 1.3 T_{\rm c}$.\cite{Lovesay95EuO,Lovesay95RbMnF}
Above $T_{\rm c}$ the correlation length $\xi$ follows the power
low $\xi\propto(T/T_{\rm c}-1)^{-\tilde{\nu}}$, where the effective critical exponent $\tilde{\nu} \simeq 0.70$
for a 3D Heisenberg magnet.\cite{Yaouanc93,Lovesay95EuO,Lovesay95RbMnF}
A fit of $\lambda$ above $T_{{\rm c}}$
to the equation $\lambda(T) = A (T/T_{\rm c}-1)^{-\tilde{\omega}}$ results in $A = 0.0030(5)$ $\mu$s$^{-1}$
and the effective dynamic exponent\cite{Keller83} ${\tilde{\omega}} = 1.06(9)$
(see the black line in Fig. \ref{fig:RlxZF}).
Thus, for $\tilde{\nu} \simeq 0.70$, we find with $\tilde{\omega}/\tilde{\nu}=1.5$ that
$\lambda \propto \xi^{1.5}$, in fair agreement with the critical behaviour of a 3D
ferromagnet ($\lambda \propto \xi^{3/2}$).

Next we discuss the $\mu$SR experiments with an electric field applied.
The temperature dependence of the mean internal magnetic field $B_1$ with or without applied electric
field $E$ is shown in Fig. \ref{fig:BintZF}(a). The temperature dependence of the internal magnetic field (i.e. the
magnetic order parameter) decreases with increasing temperature and vanishes at $T_{\rm c}$.
Analyzing the data with the power law:\cite{Pratt07}
\begin{equation}\label{eq:PowerLow}
B_1 = B_1(0)\cdot ( 1 - (T/T_c)^{\tilde{\alpha}})^{\tilde{\beta}}
\end{equation}
yields $T_{\rm c} = 57.0(1)$ K, ${\tilde{\alpha}} = 2.00(9)$,  $\tilde{\beta} = 0.39(1)$,
and $B_1(0) = 85.37(25)$ and 85.57(25)~mT
for $E = 0$ and 500/3 V/mm, respectively.
The value of the effective critical expoment\cite{Keller83}
$\tilde{\beta}$ lies quite close to the critical exponent $\beta \simeq 1/3$ expected for a 3D
magnetic  system.\cite{Jongh74} For $T\rightarrow 0$ the
temperature dependence of the internal field $B_1(0)-B_1(T)\propto T^{\tilde{\alpha}}$ is determined by the
excitation of the ground state magnetic order.\cite{Le93,Jongh74}
Figure \ref{fig:BintZF}(b) shows the difference $\Delta B_V = B_1(0{\rm V}) - B_1(500{\rm V})$ as a function of
temperature.
Obviously, the data points scatter substantially. However, a close examination shows that the data
points below 50 K are systematically shifted to negative values, except of a
 point at 10\,K [the blue open circle in Fig. \ref{fig:BintZF}(b)] which
is the first measured point and it was recorded
in the test phase of the experiment. Thus, we cannot exclude
additional systematic errors related to instrument set up for the first point.
The statistical average of all values of $\Delta B_V$ is zero within  error. By
excluding the data point at 10\,K the average value $\overline{\Delta B_V}$
below 50 K was found to be  $\simeq - 0.4(4)$\,mT.
These experiments suggest the presence of a small electric field effect on
the magnetic state of Cu$_2$OSeO$_3$.

Additional \msr experiments were performed with an applied  transverse magnetic field (TF).
The \msr spectra were found to be well described by the  asymmetry function:
\begin{align}\label{eq:TFmodel}
A_{TF}(t) = & \hat{A}_S \exp\left(-\frac{1}{2} \hat{\sigma}_{2}^2t^2\right)\cos(\gamma_\mu \hat{B}_{2} t +\phi) + \nonumber \\
+&\hat{A}_{Bg} \exp\left(-\frac{1}{2}\hat{\sigma}_{Bg}^2t^2\right)\cos(\gamma_\mu \hat{B}_{Bg} t+\phi).
\end{align}
The parameter $ \hat{A}_{S}$($ \hat{A}_{Bg}$) corresponds to $A_{S}$($A_{Bg}$) in Eq. (\ref{eq:PBasyCu}),
while $\hat{\sigma}_{2}$($\hat{\sigma}_{Bg}$) and $\hat{B}_{2}$($\hat{B}_{Bg}$) are the relaxation
rate and the mean field of the sample (Cu background), respectively.
The values of $\hat{\sigma}_{2}$ and $\sigma_2$ were found to be approximately the same,
and the Cu relaxation rate $\hat{\sigma}_{Bg}\simeq 0.25$~$\mu$s$^{-1}$ is small.
$\hat{B}_{2}$ is slightly larger than the applied field, while  $\hat{B}_{Bg}$ is close to the applied field.
In the TF \msr experiments the electric field amplitude was increased to 800/3 V/mm.
Note that the signal from the sample is well described by a single Gaussian, and that the narrow
signal in Eq. (\ref{eq:PB}) broadens.
The field dependence of $\Delta B_V' = \hat{B}_2(0{\rm V})-\hat{B}_2(800{\rm V})$ at $T=10$~K
is shown in Fig. \ref{fig:BintZF}(c).
No effect of the electric field $E$ on the internal magnetic field was found within
the precision of the experiment, although a strong field dependence of the magneto-capacitance
was reported.\cite{Bos08} The TF \msr experiment
is less precise than the ZF experiment, since the narrow signal that mainly
determines the errors becomes broader by applying a magnetic field.

\section{Conclusion}\label{sec:conclusions}

In conclusion, we performed a detailed investigation of magnetism and the magneto-electric effect in
Cu$_2$OSeO$_3$ by ZF and TF \msr.
An internal magnetic field $B_\mathrm{int}(T=0) = 85.37(25)$ mT was
detected below $T_{\rm c} = 57.0(1)$ K, consistent with a ferrimagnetic
state.\cite{Bos08} The effective critical exponent for the temperature dependence
of $B_\mathrm{int}$ was found to be  $\tilde{\beta} \simeq 0.39(1)$,
in fair agreement with the critical exponent $\beta\simeq 1/3$ expected for 3D magnetic systems.
The magnetic order parameter $B(0)-B(T) \propto T^2$ was found to exhibit a quadratic
temperature dependence for $T\rightarrow 0$.
The temperature dependence of the muon relaxation rate above $T_{\rm c}$
is well described by the relation $\lambda \propto (T/T_{\rm c}-1)^{-\tilde{\omega}}$ with
$\tilde{\omega}=1.06(9)$,  suggesting $\lambda \propto \xi^{3/2}$
($\xi$ is the magnetic correlation length) which is characteristic for 3D ferromagnets.\cite{Lovesay95EuO, Yaouanc93}
The ZF \msr measurements of the microscopic internal field distribution
with and without applied electric field $E=500/3$ V/mm
indicate a small electric field effect on the internal magnetic field:
$\Delta B_V = B_1(0{\rm V})-B_1(500{\rm V}) = - 0.4(4)$ mT.
The strong muon relaxation sets a limit on the precision of detecting a magneto-electric
effect. To improve the precision of the \msr experiment substantially higher statistics
would be needed.\\
\\
\section{Acknowledgements}

We are grateful to A. Amato and A. Raselli for their
technical support during the \msr experiments. Discussions with P.F. Meier on possible muon sites
are acknowledged. This work was performed at the Swiss Muon Source (S$\mu$S), Paul
Scherrer Institut (PSI, Switzerland). We acknowledge support
by the Swiss National Science Foundation and the NCCR {\it Materials with Novel Electronic Properties} (MaNEP).

\end{document}